\documentclass{article}
\usepackage{amsfonts}
\usepackage{amsmath}

\begin{document}

\title{Exact thermodynamics of a planar array of Ginzburg-Landau chains with
nn and nnn interaction}
\author{Victor B\^{a}rsan \\
Department of Theoretical Physics, NIPNE, Bucharest-Magurele, Romania}
\maketitle

\begin{abstract}
The exact expression of the free energy of a planar array of a
Ginzburg-Landau chains with nn and nnn interaction is obtained. The critical
behaviour of the specific heat is not qualitatively modified by taking into
account the nnn interaction.
\end{abstract}

The Ginzburg-Landau functional, sometimes called Ginzburg-Landau-Wilson
effective Hamiltonian, is of utmost interest in physics. In 2D, the real
Ginzburg-Landau fields may describe uniaxial ferromagnets \cite{R1} or
ferroelectrics \cite{R2}, ultra-thin films of hydrogen-bonded chains \cite%
{R3}, \cite{RR4} etc. Recently, the free energy of a planar array of
Ginzburg-Landau chains, with near-neighbor (nn) interaction, has been
calculated exactly and a 2D Ising-like critical behaviour of the specific
heat has been obtained \cite{R5}. This result has been extended to the 3D
case in \cite{RR5}

In this paper we shall extend the approach of \cite{R5}, taking into account
the next-near neighbor (nnn) interaction between the chains of a planar
array. The Ginzburg-Landau functional of the system is:

\begin{equation}
\mathcal{F}_{GL}[\psi ]=\sum_{j=1}^{N}\int_{0}^{L}\frac{dx}{\xi _{0}}\left[
a\psi _{j}^{2}+b\psi _{j}^{4}+c\left( \frac{d\psi _{j}}{dx}\right)
^{2}+c_{1}\left( \psi _{j+1}-\psi _{j}\right) ^{2}+c_{2}\left( \psi
_{j+2}-\psi _{j}\right) ^{2}\right]  \label{1}
\end{equation}%
where the field $\psi _{j}(x)$, on the $j-$th chain, is real, and satisfies
cyclic boundary conditions, $\psi _{j}(x)=\psi _{j+N}(x).$ The parameters $%
a,b,c,\xi _{0}$ have their usual signification, see \cite{R5} for details.

In order to obtain the thermodynamics of our system (\ref{1}), we shall
follow, as we did in \cite{R5}, the approach of Scalapino, Sears and Ferell 
\cite{R6}. According to these authors, the free energy of (\ref{1}) is
proportional to the ground state energy of a Hamiltonian associated to (\ref%
{1}) - the so-called transfer matrix Hamiltonian. In our case, the transfer
matrix Hamiltonian, associated to (\ref{1}), is:

\begin{equation}
H_{TM}=\sum_{j=1}^{N}\left[ -\frac{1}{2m}\frac{\partial ^{2}}{\partial \psi
_{j}^{2}}+a\psi _{j}^{2}+b\psi _{j}^{4}+c_{1}\left( \psi _{j+1}-\psi
_{j}\right) ^{2}+c_{2}\left( \psi _{j+2}-\psi _{j}\right) ^{2}\right]
\label{2}
\end{equation}%
It describes a 1D system of anharmonic quantum oscillators, coupled by
quadratic nn and nnn interactions; so, it presents also an intrinsic
physical interest \cite{R3}. The "mass" in (\ref{2}) is
temperature-dependent:

\begin{equation}
m=\frac{2a^{\prime }}{k_{B}^{2}T^{2}}  \label{3}
\end{equation}

It is convenient to replace (\ref{2}) with a simpler Hamiltonian:

\begin{equation}
H_{TM}=k_{B}T\tau _{n}\mathcal{H}_{TM}  \label{4}
\end{equation}%
where:

\begin{equation}
\tau _{n}=t-t_{n},\qquad t=\frac{T}{T_{MF}},\qquad t_{n}=1-\frac{2\left(
c_{1}+c_{2}\right) }{a^{\prime }},\qquad \sigma _{n}=sgn\left( t_{n}-t\right)
\label{5}
\end{equation}%
$T_{MF}$ is the "mean-field transition temperature", which, in fact, does
not correspond to any physical phase transition, but is just one of the
parameters of the Ginzburg-Landau theory, describing the temperature
dependence of $a:$

\begin{equation}
a=-a^{\prime }\tau _{n}\sigma _{n}  \label{6}
\end{equation}

In fact,

\begin{equation}
\mathcal{H}_{TM}=\sum_{j=1}^{N}\left[ -\frac{\partial ^{2}}{\partial \varphi
_{j}^{2}}-\frac{1}{2}\sigma _{n}\varphi _{j}^{2}+\lambda \varphi _{j}^{4}%
\right] +\frac{1}{2}\sum_{i,j=1}^{n}\varphi _{i}D_{ij}\varphi _{j}  \label{7}
\end{equation}%
with

\begin{equation}
\lambda =\frac{bk_{B}T_{MF}}{4a^{\prime }}\frac{t}{\tau _{n}^{3/2}}
\label{8}
\end{equation}%
and

\begin{equation}
D_{ij}=-\frac{1}{a^{\prime }\tau _{n}}\left[ c_{1}\left( \delta
_{i,j+1}+\delta _{i,j-1}\right) +c_{2}\left( \delta _{i,j+2}+\delta
_{i,j-2}\right) \right]  \label{9}
\end{equation}

The Fourier transform of the interaction potential,

\begin{equation}
D_{k}=-\frac{2}{a^{\prime }\tau _{n}}\left[ c_{1}\cos \left( ka_{0}\right)
+c_{2}\cos \left( 2ka_{0}\right) \right]  \label{10}
\end{equation}%
where $a_{0}$ is the lattice constant of the chain, plays an important role
in this approach, because the ground state of (\ref{7}) can be written as
(see \cite{R5} for details):

\begin{equation}
\mathcal{E}_{G}=\frac{1}{2}\sum_{k}\sqrt{\Omega _{0}^{2}+D_{k}}  \label{11}
\end{equation}%
where $\Omega _{0}$ is the ground state of the one-particle anharmonic
oscillator,

\begin{equation}
H=-\frac{1}{2}\frac{\partial ^{2}}{\partial x^{2}}-\frac{1}{2}\sigma
_{n}x^{2}+\lambda x^{4}  \label{12}
\end{equation}

$\Omega _{0}$ is $T-$ dependent, as a function of $\lambda ,$ according to
eq. (\ref{8}). Finally, the ground state energy of (\ref{7}) can be written
as:

\begin{equation}
\mathcal{E}_{G}=\frac{N\Omega _{0}}{2\pi }\sqrt{1+\frac{2\left(
c_{1}+c_{2}\right) }{a^{\prime }\tau _{n}\Omega _{0}^{2}}}\int_{0}^{\frac{%
\pi }{2}}\sqrt{\left( 1-k_{1}^{2}\sin ^{2}\theta \right) \left(
1-k_{2}^{2}\sin ^{2}\theta \right) }d\theta  \label{13}
\end{equation}%
where $k_{1}^{2},k_{2}^{2}$ are parameters depending on $c_{1},$ $c_{2}$ and 
$\Omega _{0}\left( T\right) .$ If

\begin{equation}
c_{1}>0,c_{2}<0,c_{1}>\left\vert c_{2}\right\vert  \label{14}
\end{equation}%
they satisfy the conditions:

\begin{equation}
k_{1}^{2}>k_{2}^{2}>0;\qquad k_{2}^{2}\rightarrow 0\qquad if\qquad
c_{2}\rightarrow 0  \label{15}
\end{equation}

So, the free energy of our system (\ref{1}) is proportional to the integral:

\begin{equation}
E\left( k_{1},k_{2}\right) =\int_{0}^{\frac{\pi }{2}}\sqrt{\left(
1-k_{1}^{2}\sin ^{2}\theta \right) \left( 1-k_{2}^{2}\sin ^{2}\theta \right) 
}d\theta  \label{16}
\end{equation}

More precisely,

\begin{equation}
F=N\Omega _{0}\frac{k_{B}T\tau _{n}}{2\pi }\sqrt{1+\frac{2\left(
c_{1}+c_{2}\right) }{a^{\prime }\tau _{n}\Omega _{0}^{2}}}E\left(
k_{1},k_{2}\right)  \label{17}
\end{equation}

The function $E\left( k_{1},k_{2}\right) $ can be considered a
generalization of the complete integral of second kind, $E(k).$ For $%
k_{2}=0, $ corresponding to the case when the nnn interaction is neglected, $%
E\left( k_{1},0\right) =E(k_{1}),$ and (\ref{17}) reduces to our previous
result, eq. (69) of \cite{R5}.

Using (252.19) and (262.17) of \cite{R7}, $E\left( k_{1},k_{2}\right) $ can
be put in the form:

\begin{equation}
E\left( k_{1},k_{2}\right) =-\frac{1}{2\alpha ^{2}\left( \alpha
^{2}-1\right) }\frac{1}{\sqrt{1-k_{2}^{2}}}\cdot  \label{18}
\end{equation}

\begin{equation*}
\left[ \alpha ^{2}E+\left( k^{2}-\alpha ^{2}\right) K+\left( 2\alpha
^{2}-\alpha ^{4}-k^{2}\right) \Pi (\alpha ^{2},k)\right]
\end{equation*}%
where

\begin{equation}
\alpha ^{2}=-\frac{k_{2}^{2}}{1-k_{2}^{2}}<0,\qquad k^{2}=\frac{%
k_{1}^{2}-k_{2}^{2}}{1-k_{2}^{2}}  \label{19}
\end{equation}%
and the conventions of \cite{R7} for elliptic integrals and functions have
been adopted.

The formulae (\ref{17}), (\ref{18}) give the exact expression of the free
energy of our system, (\ref{1}). The addition of an external "magnetic"
field $H$ can be made very simply, adding a linear term in the anharmonic
Hamiltonian (\ref{12}); in this way, $\Omega _{0}\ $will become
field-dependent, $\Omega _{0}(T,H).$ This is an important advantage of this
simpler model, compared to 2D Ising, where the addition of an external field
defies an exact solution.

Let us shortly discuss now the critical behaviour of the system. According
to (\ref{17}), (\ref{18}), the free energy is proportional to:

\begin{equation}
-k_{2}^{2}\left( 1-k_{2}^{2}\right) E(k)+k_{1}^{2}\left( 1-k_{2}^{2}\right)
K(k)+\left( k_{1}^{2}k_{2}^{2}-k_{1}^{2}-k_{2}^{2}\right) \Pi (\alpha ^{2},k)
\label{20}
\end{equation}

The singularities of this expression might appear at $k=1,$ due to the terms
in $K(k)$ and $\Pi (\alpha ^{2},k).$ In fact, the only singularity of $\Pi
(\alpha ^{2},k)$ (see \cite{R7} (904.00))

\begin{equation}
\Pi (\alpha ^{2},k)=\frac{k^{2}}{k^{2}-\alpha ^{2}}K(k)-\frac{\pi \alpha
^{2}\Lambda _{0}(\varphi ,k)}{2\sqrt{\alpha ^{2}\left( 1-\alpha ^{2}\right)
\left( \alpha ^{2}-k^{2}\right) }}  \label{21}
\end{equation}%
is due to the $K$ term, because Heuman's lambda function can be written as (%
\cite{R7} (904.01))

\begin{equation}
\Lambda _{0}(\varphi ,k)=\frac{2}{\pi }\left[ a_{0}t_{0}-\sum_{m=1}^{\infty
}a_{2m}\left( k\right) t_{2m}\left( \varphi \right) \right]  \label{22}
\end{equation}%
where

\begin{equation}
a_{0}=E,\qquad a_{2}\frac{1}{2}\left( 2K-E\right) k^{\prime 2},\qquad a_{4}=%
\frac{1}{8}\left( 4K-3E\right) k^{\prime 4},\qquad ...  \label{23}
\end{equation}

\begin{equation}
t_{0}\left( \varphi \right) =\varphi ,\qquad t_{2}\left( \varphi \right) =%
\frac{1}{2}\left( \varphi -\frac{1}{2}\sin \varphi \right) ,\qquad ...
\label{24}
\end{equation}%
and tends to

\begin{equation}
\frac{2}{\pi }\varphi E\qquad \text{if\qquad }k^{2}\rightarrow 1\qquad \text{%
or}\qquad k^{\prime 2}=1-k^{2}\rightarrow 0  \label{25}
\end{equation}

So, it is easy to see that, in (\ref{20}), for $k^{2}\rightarrow 1,$ the $K$
terms compensate each other, and the most singular contribution, coming from 
$E$ and from the second term in (\ref{16}), is proportional to

\begin{equation}
k^{\prime 2}\ln \frac{1}{k^{\prime }}  \label{26}
\end{equation}

The cancellation of the most singular terms in (\ref{20}) is quite similar
to that obtained by Fan and Wu \cite{R8} in their calculation of the
specific heat of a 2D Ising model with nnn interaction. The term (\ref{26})
gives a logarithmic singularity of the specific heat, which occurs at a
critical temperature given by the equation $k=1$ or, equivalently (see eq. (%
\ref{19})), $k_{1}^{2}=1.$ So, the presence of a nnn term does not change
qualitatively the critical behaviour of the system.

The influence of the nnn interaction on the critical behaviour was subject
of intense debate for the Ising model (as it is quite generally accepted,
the planar array of Ginzburg-Landau chains belongs to the Ising universality
class). The conclusion of several analytical approximations (\cite{R8}, \cite%
{R9}) and Monte Carlo simulations (\cite{R10}, \cite{R11}) is that the nn
Ising critical behaviour is not qualitatively modified by the nnn
interaction, at least for small values of the coupling constant $c_{2}.$

Our result agrees with these conclusions. As (\ref{3}) is also the ground
state energy of a chain of quantum anharmonic oscillators with nn and nnn
interaction, it can be also used in the study of quantum phase transitions
in this system, which, apart of being a transfer matrix Hamiltonian,
presents a certain physical interest in itself.

\end{document}